\documentclass[11pt, onecolumn]{IEEEtran} 
\pdfoutput=1
\usepackage{amsmath}
\usepackage{amssymb}
\usepackage{amsfonts}
\usepackage{amssymb,amsmath,amsthm, amsfonts}
\usepackage{epsfig}
\usepackage{subfigure}
\usepackage{calc}
\usepackage{color}
\usepackage[all]{xy}
\usepackage{bm}
\usepackage{cite}
\usepackage{tikz}
\usetikzlibrary{trees}
\usepackage{algorithmicx}
\usepackage{algpseudocode}
\usepackage{algorithm}
\usepackage{enumerate}
\usepackage{hyperref}
\usepackage{tikz}
\newcommand*\circled[1]{\tikz[baseline=(char.base)]{
            \node[shape=circle,draw,inner sep=2pt] (char) {#1};}}

\newtheorem{defn}{Definition}
\newtheorem{thm}{Theorem}[section]
\newtheorem{cor}[thm]{Corollary}
\newtheorem{prop}{Proposition}

\newtheorem{lem}[thm]{Lemma}
\newtheorem{conj}[thm]{Conjecture}
\newtheorem{constr}[thm]{Construction}
\newtheorem{note}{Remark}

\newcommand{\bit}{\begin{itemize}}
	\newcommand{\eit}{\end{itemize}}
\newcommand{\bcor}{\begin{cor}}
	\newcommand{\ecor}{\end{cor}}
\newcommand{\beq}{\begin{equation}}
\newcommand{\eeq}{\end{equation}}
\newcommand{\beqn}{\begin{equation*}}
\newcommand{\eeqn}{\end{equation*}}
\newcommand{\bea}{\begin{eqnarray}}
\newcommand{\eea}{\end{eqnarray}}
\newcommand{\bean}{\begin{eqnarray*}}
	\newcommand{\eean}{\end{eqnarray*}}
\newcommand{\ben}{\begin{enumerate}}
	\newcommand{\een}{\end{enumerate}}
\newcommand{\bdefn}{\begin{defn}}
	\newcommand{\edefn}{\end{defn}}
\newcommand{\bnote}{\begin{note}}
	\newcommand{\enote}{\end{note}}
\newcommand{\bprop}{\begin{prop}}
	\newcommand{\eprop}{\end{prop}}
\newcommand{\blem}{\begin{lem}}
	\newcommand{\elem}{\end{lem}}
\newcommand{\bthm}{\begin{thm}}
	\newcommand{\ethm}{\end{thm}}
\newcommand{\bconj}{\begin{conj}}
	\newcommand{\econj}{\end{conj}}
\newcommand{\bconstr}{\begin{constr}}
	\newcommand{\econstr}{\end{constr}}
\newcommand{\bpf}{\begin{proof}}
	\newcommand{\epf}{\end{proof}}

\newcommand{\calc}{\mbox{${\cal C}$}}
\newcommand{\calq}{\mbox{${\cal Q}$}}
\newcommand{\cmds}{\mbox{${\cal C}_{\text{\tiny MDS }}$}}

\newcommand{\fQ}{\mbox{$\mathbb{F}_Q$}}
\newcommand{\Zq}{\mbox{$\mathbb{Z}_q$}}
\newcommand{\xyin}{\mbox{$x \in \mathbb{Z}_q, y \in [t]$}}
\newcommand{\axypiz}{\mbox{$A(x,y; \pi_y(\underline{z}))$}}
\newcommand{\axyz}{\mbox{$A(x,y; \underline{z})$}}
\newcommand{\axyzexp}{\mbox{$A(x,y; z_y, \underline{z}_{\sim y})$}}
\newcommand{\axyzsw}{\mbox{$A(z_y,y; x, \underline{z}_{\sim y})$}}
\newcommand{\bxyz}{\mbox{$B(x,y; \underline{z})$}}
\newcommand{\bxypiz}{\mbox{$B(x,y; \pi_y(\underline{z}))$}}
\newcommand{\bxyzexp}{\mbox{$B(x,y; z_y, \underline{z}_{\sim y})$}}
\newcommand{\bxyzsw}{\mbox{$B(z_y,y; x, \underline{z}_{\sim y})$}}
\newcommand{\cale}{\mbox{${\cal E}$}}

\newcommand{\pz}{\mbox{$P(\underline{z})$}}	
\newcommand{\pxyz}{\mbox{$P_{(x,y)}(\underline{z})$}}	
\newcommand{\qz}{\mbox{$Q({\cal E},\underline{z})$}}

\newcommand{\uz}{\mbox{$\underline{z}$}}

\newcommand{\picz}{\mbox{$\text{Pict}({\cal E},\underline{z})$}}	
\newcommand{\hzl}{\mbox{$h(\underline{z},\ell)$}}	
\newcommand{\zyz}{\mbox{$(z_y,\underline{z}_{\sim y})$}}
\newcommand{\pizyz}{\mbox{$\pi_y(\uz) $}}	
\newcommand{\ez}{\mbox{$({\cal E},\uz)$}}
\newcommand{\cupdot}{\mathbin{\mathaccent\cdot\cup}}
\newcommand{\tlxy}{\mbox{$\theta_{\ell,(x,y)}$}}
\newcommand{\tlzyy}{\mbox{$\theta_{\ell,(z_y,y)}$}}
\newcommand{\km}{\mbox{$\mathbb{K}^m$}}
\newcommand{\zo}{\mbox{${\cal Z}_0$}}

\newcommand{\bc}{\begin{center}}
	\newcommand{\ec}{\end{center}}

\begin{document}
	
	\title{An Explicit, Coupled-Layer Construction of a High-Rate MSR Code with Low Sub-Packetization Level, Small Field Size and All-Node Repair}
\author{
  \IEEEauthorblockN{Birenjith Sasidharan, Myna Vajha, and P. Vijay Kumar} 
  
  \IEEEauthorblockA{Department of Electrical Communication Engineering, Indian Institute of Science, Bangalore.\\
    Email: \{biren, myna, vijay\}@ece.iisc.ernet.in}
}	
	\maketitle
	
	
\begin{abstract}
This paper presents an explicit construction for an $((n,k,d), (\alpha,\beta))$ regenerating code over a field $\mathbb{F}_Q$ operating at the Minimum Storage Regeneration (MSR) point.  The parameters of the MSR code can be expressed in terms of two auxiliary parameters $(q,t)$, $q \geq 2,t \geq 2$: $n=qt$, $k=q(t-1)$, $d=(n-1)$, $\alpha=q^t$ and $\beta=q^{t-1}$.   The required field size $Q$ is no larger than $n$.  The MSR code can thus be constructed to have rate $R=k/n=(t-1)/t$ as close to $1$ as desired, sub-packetization given by $r^{\frac{n}{r}}$, for $r=(n-k)$, and all code symbols can be repaired with the same minimum data download.  The construction modifies a prior construction by Sasidharan et. al. \cite{SasAgaKum} which required far larger field-size.  
A building block appearing in the construction is a scalar MDS code of block length $n$.  The code has a simple layered structure with coupling across layers, that allows both node repair and data recovery to be carried out by making multiple calls to a decoder for the scalar MDS code.   While this work was carried out independently, there is considerable overlap with a prior construction by Ye and Barg.  

It is shown here that essentially the same architecture can be employed to construct MSR codes using vector binary MDS codes as building blocks in place of scalar MDS codes.  The advantage here is that computations can now be carried out over a field of smaller size potentially even over the binary field as we demonstrate in an example.  Further, we show how the construction can be extended to handle the case of $d<(n-1)$ under a mild restriction on the choice of helper nodes.

\end{abstract}

\section{Introduction\label{sec:intro}}
	
In an $((n,k,d), (\alpha,\beta))$ regenerating code~\cite{DimGodWuWaiRam} over the finite field  $\mathbb{F}_Q$, a file of size $B$ over \fQ\ is encoded and stored across $n$ nodes in the network with each node storing $\alpha$ coded symbols.  The parameter $\alpha$ is termed as the {\em sub-packetization} level of the code. A data collector can download the data by connecting to any $k$ nodes.  In the event of node failure, node repair is accomplished by having the replacement node connect to any $d$ nodes and downloading $\beta \leq \alpha$ symbols from each node. The quantity $d\beta$ is termed the {\em repair bandwidth}.  The focus here is on exact repair, meaning that at the end of the repair process, the contents of the replacement node are identical to that of the failed node. 
	
It is well known that the file size $B$ must satisfy the upper bound (see~\cite{DimGodWuWaiRam}): 
	\bea \label{eq:cut_set_bd}
	B & \leq & \sum_{\ell =1}^{k} \min\{\alpha,(d-\ell +1)\beta\} .
	\eea
	It follows from this that $B \leq k\alpha $ and equality is possible only if $\alpha \leq (d-k+1)\beta$.   A regenerating code is said to be a Minimum Storage Regenerating (MSR) code if $B =\alpha k$ and $\alpha =(d-k+1)\beta$, since the amount $n \alpha$ of data stored for given file size $B$ is then the minimum possible. 

\subsection{Literature and Our Contribution}
While strictly speaking, the definition of an MSR code includes the requirement that all nodes be repairable with the same minimum data download, it has become customary in recent publications to refer to a code as being an MSR code even if the data download is a minimum only for the repair of systematic nodes.  We will distinguish between the two classes by referring to them as all-node-repair MSR codes and systematic-repair MSR codes respectively. 
	
There are several constructions of MSR codes to be found in the literature.  The product-matrix construction given in \cite{RasShaKum_pm}, provides MSR codes for any $2k-2 \leq d \leq n-1$.  A construction for all-node-repair MSR codes with $d = n -1 \geq 2k - 1$ is presented in~\cite{SuhRam} that builds on the systematic-repair codes constructed in \cite{ShaRasKumRam_ia}. In~\cite{PapDimCad}, high-rate MSR codes with parameters $(n, k=n-2, d=n-1)$ are constructed using Hadamard designs. In~\cite{TamWanBru}, high-rate systematic-repair MSR codes, known as zigzag codes, are constructed for $d = n-1$. This was subsequently extended to include the repair of parity nodes as well in~\cite{WanTamBru_allerton}.  In \cite{CadHuaLi}, a construction of systematic-repair MSR codes is given, that makes use of permutation matrices. In \cite{CadJafMalRamSuh}, Cadambe et al. show the existence of high-rate MSR codes for any value of $(n,k,d)$ as $\alpha$ scales to infinity.

Desirable attributes of an MSR code include an explicit construction, high-rate, low values of sub-packetization level $\alpha$ and small field size.  While zigzag codes allow arbitrarily high rates to be achieved, a level of sub-packetization that is exponential in $k$ is required.  In a subsequent paper~\cite{WanTamBru_long}, a systematic-repair MSR code having $\alpha = r^{\frac{k}{r+1}}$ is constructed. In \cite{GopTamCal}, the following lower bound on $\alpha$ is presented:
	\bea \label{eq:lb_1}
	2\log_2 \alpha (\log_{\left(\frac{r}{r-1}\right)} \alpha +1) + 1 & \geq & k.
	\eea
A second lower bound on $\alpha$, $\alpha \geq r^{\frac{k}{r}}$, can be found in \cite{TamWanBru_access_tit}, that applies to a subclass of MSR codes known as help-by-transfer (also known in the literature as access-optimal) MSR codes. For help-by-transfer MSR codes, the number of symbols transmitted as helper data over the network is equal to the number of symbols accessed at the helper nodes.   Prior to this in~\cite{CadHuaLiMeh}, the authors presented a construction of a systematic-repair MSR code that permits rates in the regime $\frac{2}{3} \leq R \leq 1$, and that has an $\alpha$ that is polynomial in $k$. In \cite{SasAgaKum}, a high-rate MSR construction for $d=n-1$ is presented that has sub-packetization level $r^{\frac{n}{r}}$ and where all nodes are repaired with minimum data download.  The construction provided was however, not explicit, and required large field size. This is extended for general $k \leq d \leq n-1$ in \cite{RawKoyVis_msr}. In \cite{GopFazVar}, the authors provide a construction for a systematic-repair MSR code for all $k \leq d \leq n-1$, but these constructions are also non-explicit and require large field size. In \cite{RavSilEtz}, explicit help-by-transfer systematic-repair MSR codes are presented with sub-packetization meeting the lower bound $\alpha \geq r^{\frac{k}{r}}$. However the constructions were limited for $r=2, 3$. 

In \cite{YeBar_1}, authors present two explicit constructions for high-rate MSR codes that allow optimal repair of $h \leq r$ simultaneous failures, by connecting to any $k \leq d \leq (n-h)$ nodes. The first construction requires sub-packetization level $\alpha = s^n$, where $s = \textsl{lcm}\{1,2,\cdots, r\}$ and a field size $\geq sn$. The second construction has the property of optimal-access, requires $\alpha = r^{n-1}$, and a field size comparable to the block-length $n$.

In the present paper, we provide an explicit help-by-transfer construction of a high-rate MSR code.
The parameters of the MSR code can be expressed in terms of two auxiliary parameters $(q,t)$, $q \geq 2,t \geq 2$: $n=qt$, $k=q(t-1)$, $d=(n-1)$, $\alpha=q^t$ and $\beta=q^{t-1}$.   The required field size $Q$ is no larger than $n$.  The MSR code can thus be constructed to have rate $R=k/n=(t-1)/t$ as close to $1$ as desired, sub-packetization given by $r^{\frac{n}{r}}$, for $r=(n-k)$, and all code symbols can be repaired with the same minimum data download.  The construction modifies a prior construction by Sasidharan et. al. \cite{SasAgaKum} which required far larger field-size. The code has a simple layered structure with coupling across layers, that allows both node repair and data recovery to be carried out by making multiple calls to a decoder for the scalar MDS code.   

In a recent paper \cite{YeBar_2}, that preceded the present work, the authors construct a high-rate MSR code with parameters identical to that of the codes presented in the present paper.    While the constructions presented in \cite{YeBar_2} and the present paper are different, and our work was carried out independently, there is significant commonality and this is discussed in greater detail in Section \ref{sec:comparison}. 

A more general viewpoint of the construction is presented in Section~\ref{sec:generic}, under which it becomes clear that the code can make use of an MDS code over an arbitrary alphabet.  In particular, the scalar MDS code to be replaced by a vector MDS code over a smaller filed, thereby potentially resulting in significant computational savings. 

Additionally, we show in Section~\ref{sec:smaller_d}, how the construction can be extended to handle the case of $q \leq d \leq n-2$ under the following mild restriction: the helper nodes must include a set of $(q-1)$ other nodes, with the choice of these other nodes being a function of the node to be repaired. 
%

\section{Description of the MSR Code}
	
\subsection{Code Parameters} 
	
Let $q\geq 2, t \geq 2$ be integers.  Let ${\mathbb{Z}_q}$ denote the set of integers modulo $q$, $[t]$ denote the set set $\{1,2,\cdots,t\}$ and $[0, q-1]$ denote the set of integers $\{0, 1, \cdots, q-1 \}$.  We describe below the construction of an $\{(n,k,d), (\alpha,\beta)\}$ high-rate MSR code over a finite field $\mathbb{F}_Q$ having parameters 
\bean
\left( n=qt, k=q(t-1), d=(n-1) \right), \ \ \left(\alpha=q^t , \beta=q^{t-1} \right)  \ \ \text{  and  } \ \ Q \leq n \ .
\eean 
Hence the code has rate $\frac{(t-1)}{t}$ and field size no larger than that of a scalar MDS code of the same block length.   

We note that through shortening, we can obtain MSR codes having $(n-\Delta,k-\Delta,d-\Delta)$ for $0 \leq \Delta \leq k-1$, starting from an MSR code with parameters $(n,k,d)$. In particular if $(n-k) \nmid n$, then we can write $n = (n-k)t - \delta = qt - \delta, \ 0 < \delta < q$. In such a case, we first construct MSR code ${\cal C}'$ with parameters $(n+\delta, k+\delta, d=n+\delta-1)$ and subsequently shorten ${\cal C}'$ to obtain a $(n,k,d=n-1)$-MSR code. 

Given a vector \uz, it will at times be found convenient to have separate access to the $y$th component, $z_y$, $y \in[t]$ of \uz.  For the reason, we define 
\bean 
\pi_y(\uz)  =  \zyz .
\eean  
We employ the notation $\pi(\cdot)$ since this is a permutation of the components of \uz.   We will write either \zyz\ or \pizyz\ depending upon whether or not we wish to draw attention to the particular component $z_y$. 
	
\subsection{The Data Cube}
	
\begin{figure}[h!]
			\begin{center}
				\subfigure[The data cube containing $( (q \times t) \times q^t ) $ symbols over the finite field \fQ. In this example, $q=4, t=5$. ]{\label{fig:cube}\includegraphics[width=2.5in]{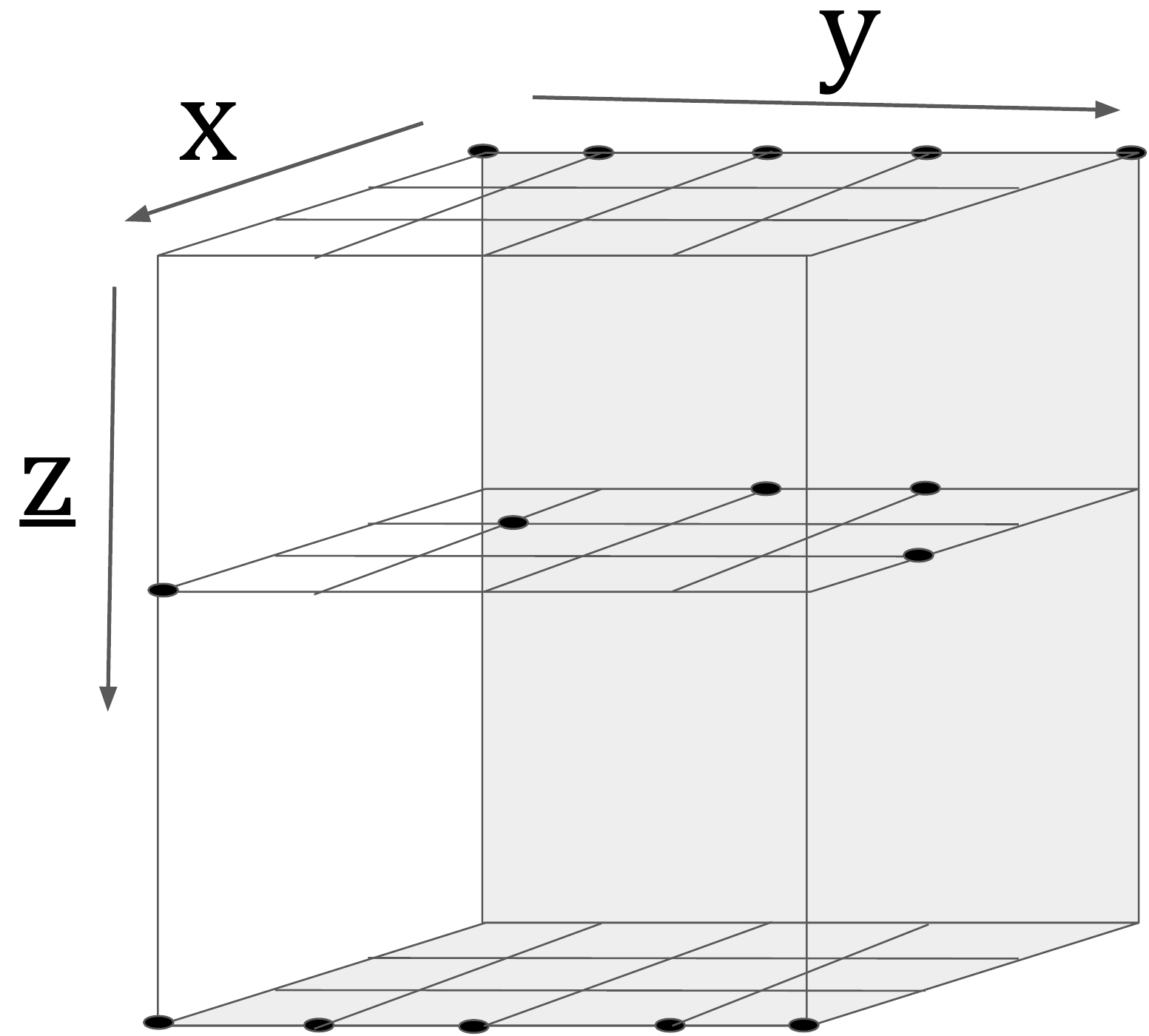}}
				\hspace{0.2in}
				\subfigure[We employ a dot notation to identify a plane. The example indicates the plane $\uz\ = (3,2,0,0,0)$.]{\label{fig:dot_rep}\includegraphics[width=2.5in]{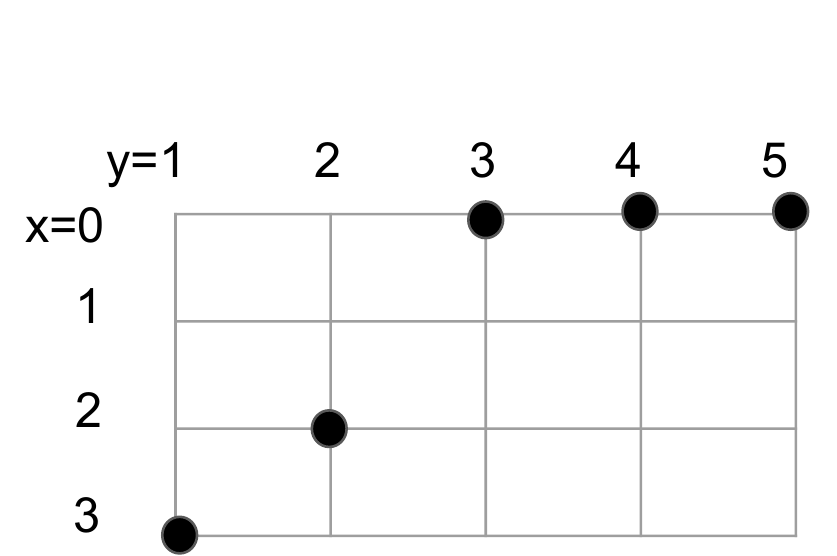}}
				\caption{Illustration of the data cube.\label{fig:repair}}
			\end{center}
		\end{figure}

	The MSR code constructed here can be described in terms of an array of symbols over \fQ\ as given below:
	\bean
	{\cal A} & = & \left\{ \axyz \mid x \in \mathbb{Z}_q, y \in [t], \underline{z} \in \mathbb{Z}_q^t \right\} .
	\eean
	This array can be depicted as a three-dimensional (3D) {\em data cube}, see Fig.~\ref{fig:cube} having dimensions $((q \times t) \times q^t)$.
In the 3D depiction of the cube, the cube appears as a collection of $q^t$ planes, with each (horizontal) plane indexed by a parameter \uz.  We associate a $(q \times t)$ $\{0,1\}$ incidence matrix $P(\underline{z})$ with each plane given by
\bean
\pxyz  & = & \left\{ \begin{array}{rl} 1 & z_y = (x-1) \\
	0 & \text{else}. \end{array} \right.
\eean
For example, if $ \uz = (1,2,3,1,0)$, we have that 
	\bean
	P(\underline{z}) & = & \left[ \begin{array}{ccccc} 0 & 0 & 0 & 0 & 1 \\ 1 & 0 & 0 & 1 & 0 \\ 0 & 1 & 0 & 0 & 0 \\ 0 & 0 & 1 & 0 & 0 \end{array} \right] .
	\eean
	
	From the point of view of the MSR code, the data cube corresponds to the data contained in a total of $n=qt$ nodes, where each node is indexed by the pair of variables:
	\bean
	\left\{ (x,y) \mid x \in \mathbb{Z}_q, y \in [t] \ \right\}  .
	\eean
	The $(x,y)$th node stores the $\alpha=q^t$ symbols 
	\bea
	C(x,y) & = & \left\{ \axyz \mid \underline{z} \in \mathbb{Z}_q^t \right\} .  \label{eq:code_and_3D_array}
	\eea  
	Thus each codeword in the MSR code is made up of the $n=qt$ vector code symbols 
    \bean 
    (C(x,y) \mid \xyin ),
    \eean in which each vector has $q^t$ components indexed by $\uz$. 
    
    Let $\Theta$ denote a parity-check matrix of an $[n,k]$-MDS code ${\cal J}$.  For example, $\Theta$ could be a Vandermonde matrix, or have form $[P \mid I]$ for $P$ a Cauchy matrix and $I$ an identity matrix, both of which can be constructed using field size $n$.   Let the rows and columns of $\Theta$ be indexed by $\ell \in [0, q-1]$ and $(x,y) \in \mathbb{Z}_q \times [t]$ respectively. We denote by \tlxy\ the  entry of $\Theta$ at the location $(\ell, (x,y))$. Let $u \in \fQ$ satisfy $u \neq 0, u^2 \neq 1$.
	
	By a slight abuse of notation, we will refer to the symbols $\axyz$ as code symbols (as opposed to calling them components of code symbols) as most of our discussion will involve the symbols $\axyz$. 
	\subsection{Parity Check Equations}
	
	The code is defined and governed by a collection of $q^{t+1}$ (linear) parity-check equations indexed by the parameter pair $(\uz, \ell)$, $\uz \in \mathbb{Z}_q^t, \ \ell \in [0,q-1]$.  The $(\uz, \ell)$th parity-check equation, which we denote by $h(\uz,\ell)$ is given by: 
	\bea
	\sum\limits_{y \in [t]}  \sum\limits_{x \in \mathbb{Z}_q}    \tlxy \axypiz \ +  u \left\{
	\sum\limits_{y \in [t]}  \sum\limits_{x \in \mathbb{Z}_q, x \neq z_y} \tlxy A(z_y,y;x,\uz_{\sim y}) \right\}
	= 0, \label{eq:pc_eqn} 
	\eea
	
	The parity-check equations can alternatively be written in the form: 
	\bea
	\sum\limits_{y \in [t]}  \sum\limits_{x \in \mathbb{Z}_q, x \neq z_y}   \tlxy \left\{ \axypiz \ +  u \axyzsw \right\} \ \  +  \sum_{y \in [t]}   \ \tlzyy \  A(z_y,y;\pizyz) = 0, \label{eq:pc_eqn_v2} 
	\eea
	for all $\ell \in [0,q-1]$ and all $ \uz \  \in \mathbb{Z}_q^t $ .  
	
	The MSR code ${\cal C}$ is then defined as the collection of all codewords 
    \bean
	  {\cal C}  =  \{ \ \left( C(x,y) \mid \xyin  \right) \ \} 
	\eean
	where each $C(x,y)$ is a $q^t$-tuple vector given by \eqref{eq:code_and_3D_array} and where the components \axypiz\ of $C(x,y)$ satisfy \eqref{eq:pc_eqn} (or equivalently \eqref{eq:pc_eqn_v2}). 

\subsection{Relation to an Earlier Construction}

In \cite{SasAgaKum}, the authors adopted a parity-check view point of an MSR code, and the code was defined through the parity-check equations: 
\bea
\label{assym_parity} \sum\limits_{y \in [t]} \sum\limits_{x \in \mathbb{Z}_q} \tlxy \axyz \ + 
\chi_{\{\ell \neq 0\}} \sum \limits_{y \in [t] } c \ A(z_y,y;\uz - \ell \underline{e}_y) = 0,
\eea
for all $\ell \in [0,q-1]$, and all $\uz \in \mathbb{Z}_q^t$. Here $\chi(\cdot)$ denotes the indicator function, and $\underline{e}_y \in \mathbb{Z}_q^t$ denotes the vector with $1$ at the $y$th coordinate, and zero everywhere else.  In that paper, the MDS property was shown to hold provided the element $c$ belonged to a sufficiently large finite field.   It is only in this sense, that the code failed to be explicit. 

The present paper arose through an attempt to find ways of guaranteeing the MDS property without resorting to a large finite field.  One such attempt led us to think of the data as being organized along $q^t$ planes, each corresponding to a different value of \uz\ .  The code symbols $\axyz$ were naturally associated with the plane \uz\ .  The parity-check equation $(\uz, \ell \neq 0)$ involved all the code symbols \axyz\ as well as code symbols associated to $t$ of the remaining planes leading to a coupling of the data belonging to the different planes.  Thus the collection of parity check equations associated with plane \uz\ letting all possible $\ell$, $0 \leq \ell \leq (q-1)$, involves symbols from a set ${\cal P}_{\underline{z}}$ of $(q-1)t$ different planes. The problem of data collection could then be viewed as a process in which the data belonging to various planes was recovered sequentially.   Attempts at proving the MDS property under this approach, led us to the need to establish the non-singularity of matrices having a block-matrix structure in which the simplest instance of these took on the form:
\bean
D & = & \left[ \begin{array}{ccc|ccc}
				1 & 1 & 1 &  & & \\
				\theta_1 & \theta_2 & \theta_3 & c & & \\
				\theta_1^2 & \theta_2^2 & \theta_3^2 &  & & \\
				&  & & 1 & 1 & 1  \\
				&  & & \theta_1 & \theta_2 & \theta_3  \\
				& c &  & \theta_1^2 & \theta_2^2 & \theta_3^2 \end{array} \right].
\eean
Proving the invertibility of such matrices turned out to be quite challenging, particularly as the number of blocks increased.   Various attempts to modify the matrix above in manner that would enable easier proof of non-singularity were attempted.  One such attempt in which the matrix above was replaced by the matrix 
\bean
D' & = & \left[ \begin{array}{ccc|ccc}
				1 & 1 & 1 & u & & \\
				\theta_1 & \theta_2 & \theta_3 & u\theta_2 & & \\
				\theta_1^2 & \theta_2^2 & \theta_3^2 & u\theta_2^2 & & \\
				& u & & 1 & 1 & 1  \\
				& u\theta_1 & & \theta_1 & \theta_2 & \theta_3  \\
				& u\theta_1^2 &  & \theta_1^2 & \theta_2^2 & \theta_3^2 \end{array} \right]
\eean
proved successful and led to the present construction.  This increases the amount of coupling and it can be verified in the present construction, that the parity check equation $(\uz, \ell)$ involves symbols from the same set ${\cal P}_{\underline{z}}$ of $(q-1)t$ planes for {\em every} $0 \leq \ell \leq (q-1)$.

\subsection{Transformed Code Symbols \label{sec:transform}} 
	
	Let us introduce the variables
	\beqn
	\bxypiz =  \left\{ \begin{array}{rl} \axypiz + u \axyzsw,  x \neq z_y \\
		\axypiz,   x = z_y \end{array} \right. .
	\eeqn
	Interchanging variables $x,z_y$, for $x \neq z_y$, we obtain  
	\bean
	\bxyzsw & = & u \axypiz +  \axyzsw .
	\eean
	We will refer to the symbol pair $\axyzexp,A(z_y,y;x,\underline{z}_{\sim y})$ as {\em companion} terms.  Similarly with the pair $\bxyzexp,B(z_y,y;x,\underline{z}_{\sim y})$. 
	For $(x,z_y)$, $x\neq z_y$, this gives us an invertible transformation between the paired companions:
	\beq
	\left[ \begin{array}{c} \bxyzexp \\ \bxyzsw \end{array} \right]  =  
	\left[ \begin{array}{cc} 1 &  u \\ u & 1 \end{array} \right] 
	\left[ \begin{array}{c} 
		\axyzexp \\ \axyzsw
	\end{array} \right] . \label{eq:transform_1}
	\eeq 
	This forces $u^2 \ne 1$. We will use ${\cal L}$ and ${\cal L}^{-1}$ to refer to the linear transformation and its inverse. In terms of this notation, we would then have that:
	\beqn
	{\cal L}(A(x,y;\uz), \axyzsw) = (B(x,y;\uz), \bxyzsw)
	\eeqn

	When expressed in terms of the array \bxypiz, the parity equations take on the simplified form:
	\beq \label{eq:bcode}
	\sum\limits_{x \in \mathbb{Z}_q} \sum\limits_{y \in [t]}   \tlxy \bxyzexp =  0, \ \underline{z} \in \mathbb{Z}_q^t, \ \ell \in [0,q-1].  
	\eeq
	
	\begin{figure}[h!]
		\begin{center}
				\subfigure[The symbols $A(1,1;3,\uz_{\sim 1}), A(3,1;1,\uz_{\sim 1})$ marked as $\bigtriangleup$ are companion terms that get coupled to form $B(1,1;3,\uz_{\sim 1}), B(3,1;1,\uz_{\sim 1})$. where $\uz_{\sim 1} = (0,0,0,0)$.]{\label{fig:symbol_coupling}\includegraphics[width=2in]{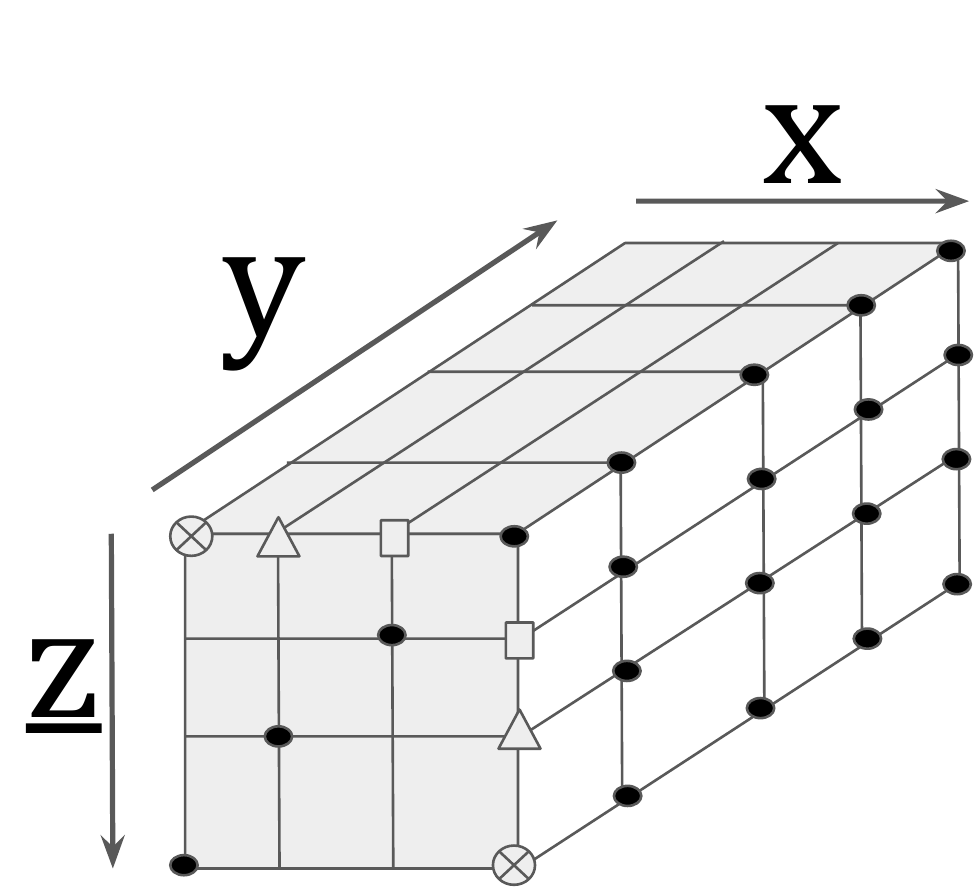}}
				\hspace{0.9in}
				\subfigure[Entire data cube is shown for the case of $q=2,t=3$. We show three pairs of companion terms where terms in each pair are connected by dotted lines.]{\label{fig:symbol_coupling_2}\includegraphics[width=1.1in]{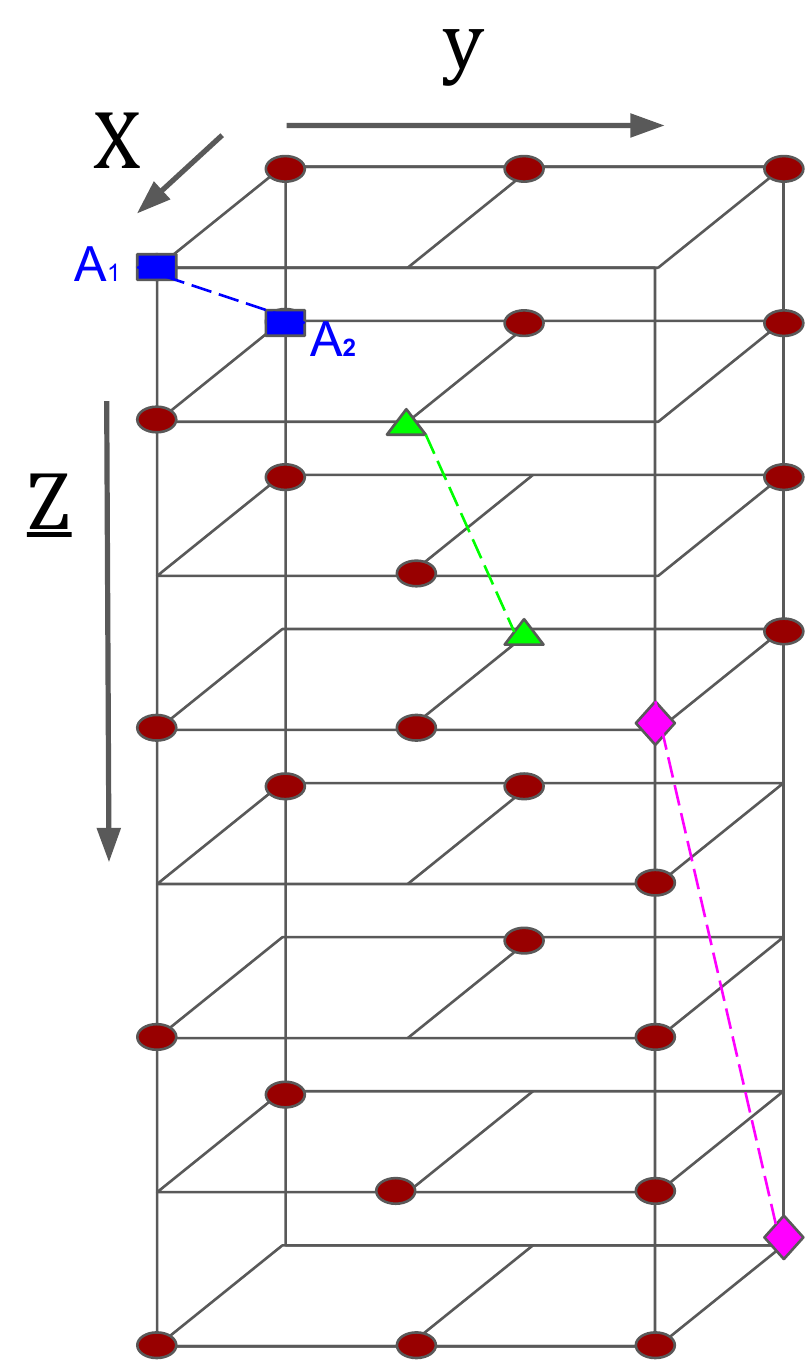}}
				\caption{Illustration of the coupling between comapnion terms.\label{fig:coupling}}
		\end{center}	
	\end{figure}

    \subsection{Interpretation in Terms of a Layered MDS Code with Uncoupled Layers}
    
    If in the definition of the code ${\cal C}$ we were to replace each code symbol \axyz\ by \bxyz\, where $\{\axyz\}$ and $\{\bxyz\}$ are related by \eqref{eq:transform_1}, then we will be lead to a second vector code ${\cal D}$, also of block length $n=qt$.  Each vector code symbol $D(x,y)$ in ${\cal D}$ would then be a vector having the $q^t$ components
    \bean
    D(x,y) & = & (\bxyz\ , \uz \in \mathbb{Z}_q^t). 
    \eean
    The code ${\cal D}$ can be verified to be a code obtained by layering $q^t$ MDS codes, each of block length $n$ and rate $\frac{(t-1)}{t}$.  Such a layered code can trivially be verified to have the data collection property required of an MSR code.  One simply collects data from each layer by recovering the MDS code from $k$ of the symbols in the layer.  However, such a code would fail to meet the minimum data download requirement of an MSR code.  In relation to code ${\cal D}$, the code ${\cal C}$ can be viewed as an modification of the code ${\cal D}$ in which coupling between code symbols across layers is introduced to facilitate node repair.  For this reason, we will refer to ${\cal D}$ as the {\em decoupled} code. 
   	As we will see subsequently, data collection in the case of code ${\cal C}$ can be accomplished in a manner very similar to that one would have employed in the case of code ${\cal D}$, except for two differences:
   	\bit
   	\item one has to carefully select the order of decoding symbol.
   	\item while decoding a layer, one will repeatedly invoke the inverse transformation ${\cal L}^{-1}$ to undo the coupling across layers. 
   	\eit  
   	This will become clear in the sequel.
    
	\subsection{The Associated Scalar Code} 
	
	This subsection can be skipped on a  first reading without loss of continuity. Let the collection of $(q \times t \times q^t)$ symbols \axyz\ form a scalar code that satisfies the same collection $\{\hzl \mid \uz \in \Zq^t, \ell \in \Zq\}$ of parity-check equations given by \eqref{eq:pc_eqn}.  Let $H_{\text{scalar}}$ be the associated parity-check matrix whose coefficient at row indexed by $(\uz, \ell)$ and column indexed by $(a,b,\underline{z}')$ is given by 
	\bean
	\phi( (\uz, \ell),(a,b ;\uz') ).
	\eean
	The $(\uz\ , \ell)$th parity-check equation would then read as :
	\bean
	\sum_{a \in \mathbb{Z}_q,  \ b \in [t], \  \uz' \in \mathbb{Z}_q^t} \phi \left( (\uz , \ell),(a,b; \underline{z}') \right) A(a,b; \uz') & = & 0. 
	\eean
	In comparison, we have that 
	\bea
	\sum\limits_{y \in [t]} \sum\limits_{x \in \mathbb{Z}_q}    \theta_{\ell, (x,y)} \axyzexp  \ + \ u \sum_{y \in [t]}  \sum_{x \in \mathbb{Z}_q, \ x \neq z_y}   \ \theta_{\ell,(x, y)} \  A(z_y,y;x,\underline{z}_{\sim y}) = 0, \label{eq:pc_eqn_v3} 
	\eea
	By comparing terms we conclude that 
	\bean
	\phi \left( (\uz\ , \ell),(a,b;z'_b, \uz'_{\sim b}) \right)  & = & 
	\left\{ \begin{array}{rl}
		\theta_{\ell,(a,b)}  & \uz' = \uz \\
		u \theta_{\ell,(z'_b,b)}, & a=z_b \ne z_b',\ \uz'_{\sim b} =\underline{z}_{\sim b} \\
		0 &  \text{else} .
	\end{array} \right.
	\eean
	Each row has Hamming weight (i.e., number of nonzero entries) equal to $qt+t(q-1)=t(2q-1)$.  Let us consider a code symbol $A(a, b; z'_b, \uz'_{\sim b})$. It is clear to see that the symbol occurs in $q$ parity-check equations $\{ \ h(\uz',\ell) \mid \ell \in [0, q-1] \ \}$. Suppose the symbol $A(a,b; z'_b, \uz'_{\sim b})$ participates in a parity-check equation $h(\underline{z}, \ell)$ where $\underline{z} \ne \uz'$. This would imply that $a = z_b \ne z'_b $ and $\underline{z}_{\sim b} = \uz'_{\sim b}$. So whenever $z'_b = a$,  the symbol $A(a, b; a, \uz'_{\sim b})$ appears only in $q$ parity-check equations. On the other hand when $z'_b \ne a$, $\uz$ is uniquely determined by $z_b = a, \uz_{\sim b} = \uz'_{\sim b}$. Hence the symbol participates in additional $q$ parity-check equations $\{ \ h(\uz, \ell) \mid z_b = a, \uz_{\sim b} = \uz'_{\sim b}, \ell \in [0,q-1] \ \}$. Therefore the columns have Hamming weight equal to either $q$ or $2q$.  
	

\section{Verifying the Repair Property of an MSR Code \label{sec:rep}}
	
Let us assume that the node having node index $(x_0,y_0)$ has failed.  The goal then is to recover the values of 
\bean
C(x_0,y_0) & = & \{ A(x_0,y_0; \uz) \mid \ \uz \in \mathbb{Z}_q^t \},
\eean
for all values of $\underline{z}$ by downloading at most $\beta=q^{t-1}$ symbols from each of the remaining $d=(n-1)$ nodes.  Each of the $d=qt-1$ helper nodes corresponds to a distinct pair $(x,y) \in (\mathbb{Z}_q \times [t])$, $(x,y) \neq (x_0,y_0)$.  We will show that repair of node $(x_0,y_0)$ can be accomplished by downloading only the $\beta=q^{t-1}$ symbols $ \{\axyz \mid z_{y_0} = x_0\} $,	from helper node having index $(x,y)$.  

	\begin{figure}[h!]
		\begin{center}
			\subfigure[The gray vertical pillar on the extreme left identifies the failed node $(3,1)$. The symbols belonging to planes $(z_1 = 3, \underline{b}_{\sim y_0})$ for every $\underline{b}_{\sim y_0}$, identified in gray, are transmitted as helper data. ]{\label{fig:repair_cube}\includegraphics[width=3.0in]{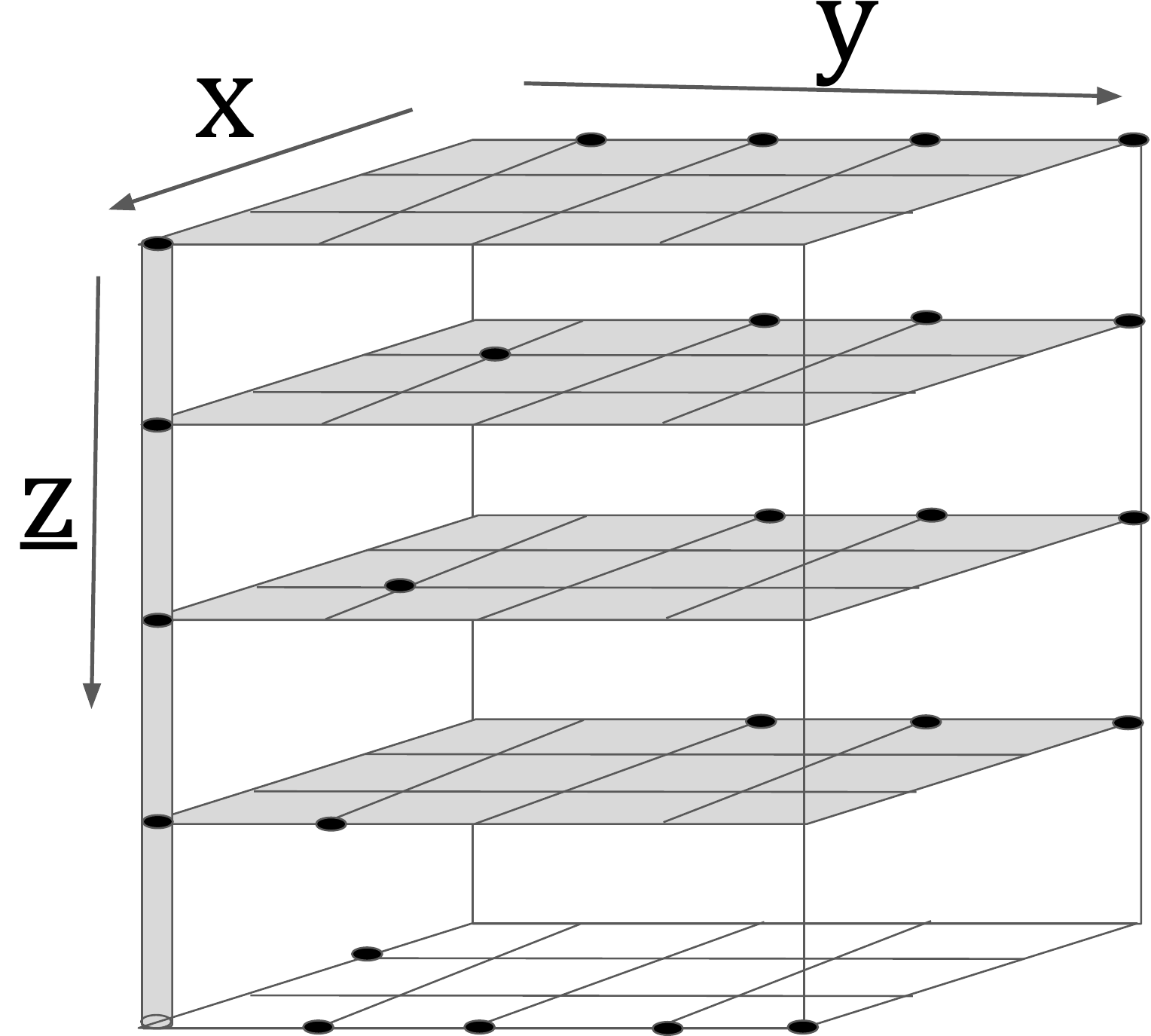}}
			\hspace{0.5in}
			\subfigure[In this zoomed-in view, we show repair of $q=4$ circled symbols at the node $(3,1)$, using the helper data from the plane $(3,0,0,0,0)$ (identified in gray).]{\label{fig:repair_cube_2}\includegraphics[width=2.2in]{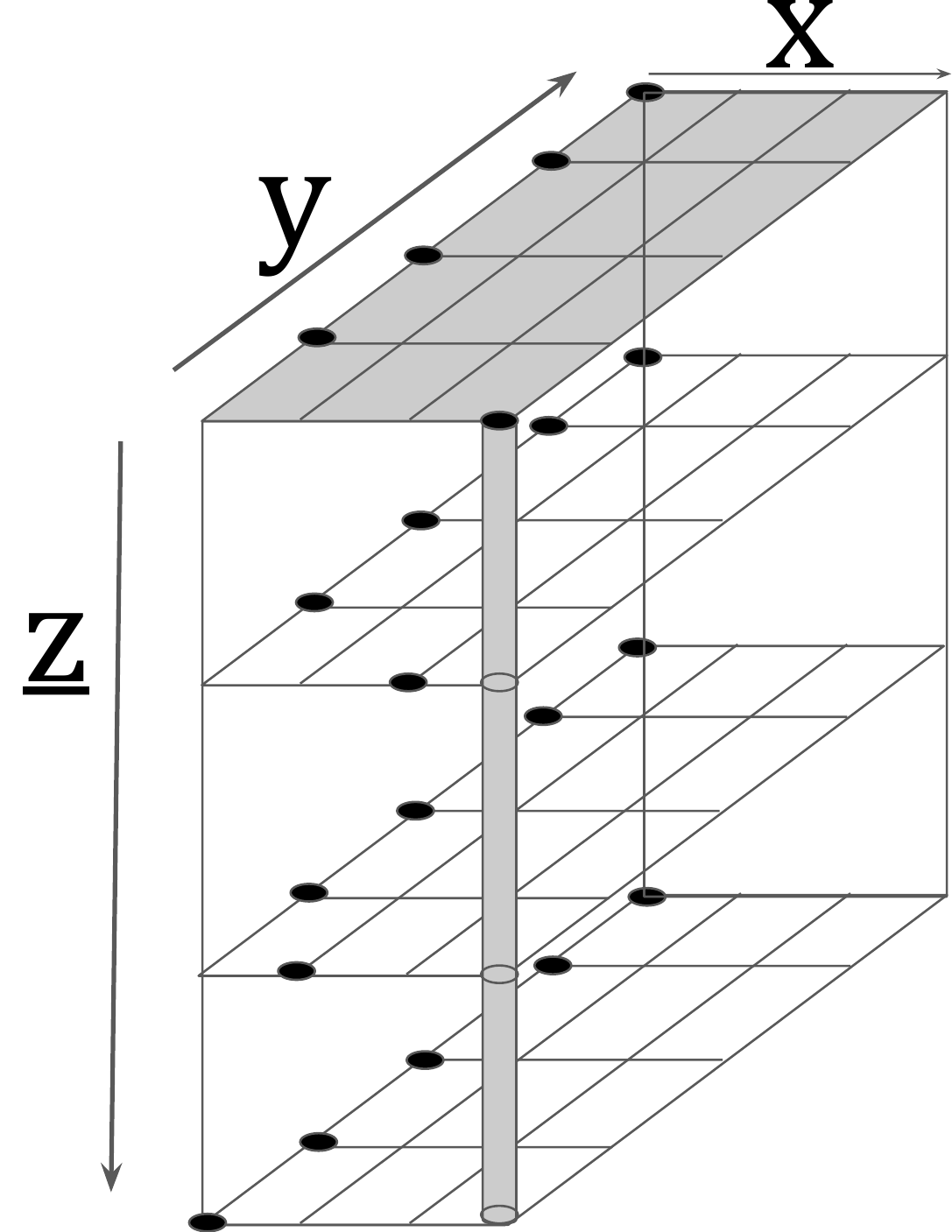}}
			\caption{Illustration of node repair using data cube.\label{fig:repair}}
		\end{center}
	\end{figure}
In the repair process, let us use $\kappa_{*}$ (mnemonic for known) to denote any function of the $(qt-1)q^{t-1}$ symbols
\bean
\{A(x,y;\zyz) \mid (x,y) \neq (x_0,y_0), z_{y_0}=x_0 \},
\eean
downloaded for repair of node $(x_0,y_0)$.  
Consider the parity-check equations $h(\uz,\ell)$ for all $\ell \in [0,q-1]$, associated to $(z_{y_0}, \underline{z}_{\sim y_0})=(x_0,\underline{b}_{\sim y_0})$ for fixed $\underline{b}_{\sim y_0} \in \mathbb{Z}_q^{t-1}$.  These can be expressed in the form
	\bea
	\theta_{\ell,(x_0,y_0)}  A(x_0,y_0;x_0, \underline{b}_{\sim y_0}) \ +  \sum \limits_{x \in \mathbb{Z}_q, x \neq x_0  } u  \theta_{\ell,(x, y_0)} A(x_0,y_0;x,\underline{b}_{\sim y_0}) = \kappa_{*} .  \label{eq:node_repair} 
	\eea
	By the MDS property of $\Theta$ and by the choice of $u \neq 0$, these $q$ equations can be solved to determine the $q$ unknown code symbols 
	\bean
	A(x_0,y_0;x_0, \underline{b}_{\sim y_0}) \cup \{ A(x_0,y_0;x, \underline{b}_{\sim y_0}) \mid x \in \mathbb{Z}_q, x \neq x_0 \}.
	\eean
	By repeating this process for all possible $\underline{b}_{\sim y_0} \in \mathbb{Z}_q^{t-1}$, we will have recovered all the code symbols of node $(x_0, y_0)$.
 
\section{Notation to Handle Data Collection} 

\subsection{Erasure Matrix} Let ${\cal E} = \{  (x_i, y_i) \in \mathbb{Z}_q \times [t] \mid 1 \leq i \leq q\}$ denote the location of the $q$ erased nodes.  
We associate with this erasure pattern, the $(q \times t)$ {\em erasure} matrix 
\bean
E_{(x,y)}({\cal E}) & = & \left\{ \begin{array}{rl} 1 & (x,y)=(x_i,y_i)  \\
	0 & \text{else}. \end{array} \right.
\eean
For example, when $(q=4,t=5)$ and ${\cal E} = \{ (0,2), (1,2),(2,2),(2,4) \}$ we have that 
\bea
E({\cal E}) & = & \left[ \begin{array}{ccccc} 0 & 1 & 0 & 0 & 0 \\ 0 & 1 & 0 & 0 & 0 \\ 0 & 1 & 0 & 1 & 0 \\ 0 & 0 & 0 & 0 & 0 \end{array} \right]. \label{eq:example_erasure} 
\eea

\subsection{Intersection Score of an Erasure Pattern on a Plane}  Given a plane \uz\ $\in \mathbb{Z}_q^t$ associated to matrix \pz\ and an erasure pattern ${\cal E}$, we define the {\em intersection score} $\sigma\ez$ to be given by
\bea
\sigma\ez & = & \mid \left\{ y \in [t] \mid (z_y,y) \in \cale \right\} \mid. \label{eq:score_def}
\eea
We define $\sigma_{\max}(\cale)=\max \{\sigma(\cale,\uz) \mid \uz \in \mathbb{Z}_q^t\}$. 
 
This quantity can also be defined as the Hamming weight of the $\{0,1\}$ matrix $Q\ez$ that is the Schur (component-wise) product of the matrices \pz\ and $E({\cal E})$: 
\bean
\qz\ & = & \pz\ \odot E({\cal E}),
\eean
and this explains the nomenclature. 
In the case of the example matrices given above, we have 
\bean
\qz\ & = & \left[ \begin{array}{ccccc} 0 & 0 & 0 & 0 & 1 \\ 1 & 0 & 0 & 1 & 0 \\ 0 & 1 & 0 & 0 & 0 \\ 0 & 0 & 1 & 0 & 0 \end{array} \right] \odot  \left[ \begin{array}{ccccc} 0 & 1 & 0 & 0 & 0 \\ 0 & 1 & 0 & 0 & 0 \\ 0 & 1 & 0 & 1 & 0 \\ 0 & 0 & 0 & 0 & 0 \end{array} \right] \ \\ &=& \ 
\left[ \begin{array}{ccccc} 0 & 0 & 0 & 0 & 0 \\ 0 & 0 & 0 & 0 & 0 \\ 0 & 1 & 0 & 0 & 0 \\ 0 & 0 & 0 & 0 & 0 \end{array} \right].
\eean
Hence we have for this example $\sigma \ez = 1$, $\sigma_{\max}(\cale) = 2$.

\subsection{Pictorial Representation} We provide below a pictorial representation $\text{Pict}({\cal E},\uz )$ which captures in one figure, both the erasure pattern ${\cal E}$ as well as the plane \uz\ under examination.
\beqn
\pz = \left[ \begin{array}{ccccc} 0 & 0 & 0 & 0 & 1 \\ 1 & 0 & 0 & 1 & 0 \\ 0 & 1 & 0 & 0 & 0 \\ 0 & 0 & 1 & 0 & 0 \end{array} \right], \ \ \ 
{\cal E} =\left[ \begin{array}{ccccc} 0 & 1 & 0 & 0 & 0 \\ 0 & 1 & 0 & 0 & 0 \\ 0 & 1 & 0 & 1 & 0 \\ 0 & 0 & 0 & 0 & 0 \end{array} \right] \eeqn
\beqn
\text{Pict}({\cal E},\underline{z}) \ =\ \left[ \begin{array}{ccccc} 0 & \circled{0} & 0 & 0 & 1 \\ 1 & \circled{0} & 0 & 1 & 0 \\ 0 & {\color{red} \circled{1}} & 0 & \circled{0} & 0 \\ 0 & 0 & 1 & 0 & 0 \end{array} \right] .
\eeqn
The locations of the circled $1$s (shown in red) in \picz\  identify the intersection `points' and the number of them is the intersection score, and hence $\sigma \ez = 1$ in the present example.

\section{Sequential Decoding Approach to Data Collection \label{sec:seq_dec}}  

The data collection property of the MSR code requires that the entire data be recoverable by connecting to any $k$ nodes. Equivalently, we should be able to recover from any $(n-k)=q$ node erasures. In this section, we will show how such a recovery can take place by providing a sequential decoding algorithm that proceeds in multiple rounds. Given erasure pattern \cale\, in the $s$th round $0 \leq s \leq \sigma_{\max}(\cale)$, we will decode the erased symbols in planes having intersection score $\sigma=s$ and make use of symbols decoded in prior rounds.   A pseudocode for the algorithm appears in Algorithm~\ref{alg:seq_dec} in which {\bf S-MDS-DEC} is a reference to a decoder for the scalar MDS code ${\cal J}$.

\subsection{Restricting the Parity-Check Equations to Just the Erased Symbols} 

Let \cale\ be a fixed erasure pattern.  The aim is to recover the erased code symbols, i.e., recover the values 
\bean
\{ A(x_i,y_i;\uz) \mid (x_i, y_i) \in \cale, \uz\ \in \mathbb{Z}_q^t \} . 
\eean
Since all the non-erased symbols are available to the decoder, we can equivalently rewrite the parity-check equations in the form:
\bean
\sum\limits_{\small \begin{array}{c} y \in [t], \ x \in \Zq, \\ (x,y) \in \cale \end{array}}     \tlxy \axyzexp +  u \left\{ 
\sum\limits_{\small \begin{array}{c} y \in [t], \ x \in \Zq, \\ x \neq z_y, (z_y,y) \in \cale \end{array}} 
\tlxy A(z_y,y;x,\uz_{\sim y}) \right\} 
& = & \kappa_{*},
\eean
where $\kappa_{*}$ is generic notion for a known element in the finite field $\mathbb{F}_Q$ that can be determined from the non-erased code symbols.  Suppressing notation for the well-known ranges over which the variables $x,y$ vary, we can more simply write this in the form
\bea
\sum\limits_{(x,y) \in \cale }     \tlxy \axyzexp +  u \left\{ 
\sum\limits_{(z_y,y) \in \cale,  \ x \neq z_y } 
\tlxy A(z_y,y;x,\underline{z}_{\sim y}) \right\} 
& = & \kappa_{*}. \label{eq:focused_pc_eqn} 
\eea
\subsection{Case of Zero Intersection Score  }

Consider a plane \uz\ which is such that $\sigma\ez=0$.  In this case, the plane \uz\ and the erasure pattern \cale\ are such that there is no $y \in [t]$ such that $(z_y,y) \in \cale$.  Hence the second summation term in \eqref{eq:focused_pc_eqn} is vacuous and so \eqref{eq:focused_pc_eqn} reduces simply to 
\bea
\sum\limits_{\small (x,y) \in {\cal E}}     \tlxy \axyzexp 
& = & \kappa_{*}. \label{eq:wh_eq_0} 
\eea
This is a set of $q$ equations in $q$ unknowns with an invertible coefficient matrix by the choice of $\Theta$. Hence the unknowns can be solved for. 

\subsection{Case of Intersection Score $\sigma>0$ }

Here we will show how one can inductively recover code symbols corresponding to planes \uz\ having intersection score $\sigma>0$, given that symbols in planes $\uz'$ with $\sigma({\cal E}, \uz') < \sigma$ have already been recovered.  We have already carried out recovery of code symbols in planes with intersection score $0$, settling the first step of the induction. 

Let an erasure pattern \cale\ and a plane \uz\ be fixed.  Let us define 
\bean
\cale_{0,\underline{z}} & = & \left\{ (x,y) \in \cale \mid x=z_y \right\}, \\
\cale_{1,\underline{z}} & = & \left\{ (x,y) \in \cale \mid (z_y,y) \notin \cale \text{ hence $x \neq z_y$} \right\}, \\
\cale_{2,\underline{z}} & = & \left\{ (x,y) \in \cale \mid (z_y,y) \in \cale, \ x \neq z_y \right\}    .
\eean
Clearly, this represents a partition of the set of $q$ erasures into disjoint subsets: 
\bean
\cale & = & \cale_{0,\underline{z}} \cupdot \cale_{1,\underline{z}} \cupdot \cale_{2,\underline{z}}
\eean
for any plane $\uz$.
With respect to the parity-check equations in \eqref{eq:focused_pc_eqn} restricted to the erased symbols, the second summation term in \eqref{eq:focused_pc_eqn} involves code symbols 
\bean
A(z_y,y;x,\underline{z}_{\sim y})  \text{ with $(x,y,\uz)$ satisfying } 	(z_y,y) \in \cale , \ x \neq z_y .
\eean
Consider the intersection score $ \sigma({\cal E},(x,\underline{z}_{\sim y}))$ of the plane $(x,\underline{z}_{\sim y})$. Since $(z_y,y) \in \cale$, it follows that
\bean
(x,y) \notin \cale \ \Rightarrow \sigma({\cal E},(x,\underline{z}_{\sim y})) < \sigma \ez . 
\eean	
But this implies that the corresponding symbols $A(z_y,y;x,\underline{z}_{\sim y})$ have already been recovered in a prior step of the sequential decoding process. Hence we can move such terms to the right hand side and absorb them into the symbol $\kappa_{*}$ which represents the accumulated past knowledge of previously recovered code symbols. This means that we can rewrite \eqref{eq:focused_pc_eqn} in the form
\bea
\sum\limits_{(x,y) \in {\cal E} }     \tlxy \axyzexp \ + \ u \left\{  
\sum\limits_{\small \begin{array}{c} (x,y) \in {\cal E} \\ (z_y,y) \in {\cal E},  \ x \neq z_y \end{array} } 
  \tlxy A(z_y,y;x,\underline{z}_{\sim y}) \right\}
 &= & \kappa_{*}. \label{eq:focused_pc_eqn_v2} 
\eea
where in the second summation, we have in view of the argument above, added the further requirement that $(x,y) \in \cale$.    But $(z_y,y) \in \cale, \ x \neq z_y$ implies that $(x,y) \in \cale_{2,\underline{z}}$.  Hence we can change the restriction in the second summation from
\bean
(x,y) \in \cale, (z_y,y) \in \cale,  \ x \neq z_y, 
\eean
simply to $(x,y) \in \cale_{2,\underline{z}}$. This allows us to rewrite \eqref{eq:focused_pc_eqn_v2} in the form
\bea
\sum\limits_{(x,y) \in {\cal E} }     \tlxy \axyzexp \ + \ u \left\{
\sum\limits_{(x,y) \in {\cal E}_{2,\underline{z}} } 
 \tlxy A(z_y,y;x,\underline{z}_{\sim y}) \right\}   
 &=&  \kappa_{*}. \label{eq:focused_pc_eqn_v3} 
\eea
Combining terms from both summations that correspond to $(x,y) \in \cale_{2,\underline{z}}$, we obtain 
\bean
\sum_{(x,y) \in  {\cal E}_{0,\underline{z}} \cupdot {\cal E}_{1,\underline{z}}} \tlxy A(x,y;\underline{z}) \ +  \ 
\sum_{(x,y) \in {\cal E}_{2,\underline{z}}} \tlxy \left\{  \axypiz \ + \ u A(z_y,y;x,\underline{z}_{\sim y})\right\}   &= & \kappa_{*},
\eean
i.e., 
\bea
\sum_{\substack{(x,y) \in \\ {\cal E}_{0,\underline{z}} \cupdot {\cal E}_{1,\underline{z}}}} \tlxy A(x,y;\uz) \ + 
\sum_{(x,y) \in {\cal E}_{2,\underline{z}}} \tlxy B(x,y;\uz)   &=&  \kappa_{*}. \label{eq:focused_pc_eqn_v4}
\eea
But this is now a collection of $q$ equations in $q$ unknowns and can hence be solved to obtain 
\bean
\{ A(x,y;\uz) \mid (x,y) \in  \cale_{0,\underline{z}} \cupdot \cale_{1,\underline{z}}\} \ \bigcup \ 	\{ B(x,y;\uz) \mid (x,y) \in  \cale_{2,\underline{z}} \}.
\eean
Consider the terms $\{ B(x,y;\uz) \mid (x,y) \in  \cale_{2,\underline{z}} \}$.  Our aim was to recover instead the terms $\{ A(x,y;\uz) \mid (x,y) \in  \cale_{2,\underline{z}} \}$.  This can be accomplished if we are also able to compute the companion terms 
\bean
\{ B(z_y,y;x, \underline{z}_{\sim y}) \mid (x,y) \in  \cale_{2,\underline{z}} \}
\eean
which will allow us to employ the transformation ${\cal L}^{-1}$ to recover the terms $\{ A(x,y;\uz) \mid (x,y) \in  \cale_{2,\underline{z}} \}$.

Towards this end, consider a specific term $B(x_0,y_0;z_{y_0},\underline{z}_{\sim y_0})$ whose companion $B(z_{y_0},y_0;x_0,\underline{z}_{\sim y_0})$ terms we wish to compute. Consider the recovery of the symbols in the plane $\uz' = (x_0,\underline{z}_{\sim y_0})$, under the same erasure pattern \cale.  With respect to the plane $\uz'$, we can partition $\cale$ as $\cale  =  \cale_{0,\underline{z}'} \cupdot \cale_{1,\underline{z}'} \cupdot \cale_{2,\underline{z}'}$. By definition of intersection score in \eqref{eq:score_def}, it is clear that $\sigma({\cale, \uz}) = |{\cal E}_0|$. It can be seen that
\bean
(z_{y_0},y_0) \in \cale_{0,\underline{z}} \ , \  (x_0,y_0) \in \cale_{0,\underline{z}'}.
\eean
and any $(z_y,y) \in {\cale}_{0,\underline{z}}$ such that $y \neq y_0$ will be an element of ${\cale}_{0,\underline{z}'}$ as well. Hence both ${\cale}_{0,\underline{z}}$ and ${\cal E}_{0,\underline{z}'}$ have the same size. Therefore, the planes \uz\ and $(x_0,\underline{z}_{\sim y_0})$ have precisely the same intersection score,  and thus will be decoded in the same round, leading to the recovery of the symbols 
\bean
\{ B(x,y;x_0,\underline{z}_{\sim y_0})) \mid (x,y) \in  \cale_{2,\underline{z}'} \}.
\eean
This includes the specific companion term that we are looking for, namely \bean B(z_{y_0},y_0;x_0,\underline{z}_{\sim y_0}) \eean
as $(z_{y_0}, y_0) \in \cale_{2,\underline{z}'}$. With this, we are thus able to decode all symbols \bean A(x,y;\underline{z}) \eean 
as desired. In summary, during the $s$th round, we first recover a mixture of symbols 
\bean 
A(x,y;\uz) \mid (x,y) \in  \cale_{0,\underline{z}} \cupdot \cale_{1,\underline{z}}\} \ \bigcup \ 	\{ B(x,y;\uz) \mid (x,y) \in  \cale_{2,\underline{z}} \} 
\eean
for each plane \uz\ .  At the end of the round, we will have recovered for every term $B(x,y;z_y,\underline{z}_{\sim y})$, its companion term $B(z_{y},y;x,\underline{z}_{\sim y})$ and this allows us to recover the desired symbols $A(x,y;z_y,\underline{z}_{\sim y})$ and $A(z_{y},y;x,\underline{z}_{\sim y})$ . 

\begin{algorithm}[h!]
	\caption{Sequential Decoding \label{alg:seq_dec}}
	\begin{algorithmic}[1]
		\State Input: \cale.
		\State Compute $\sigma_{\max}(\cale)$, set $s=0$.
		\State Assign intersection scores $\sigma(\cale,\uz)$ to all planes $\uz \in \mathbb{Z}_q^t$.
		\While {( $s \le \sigma_{\max}(\cale)$ ) }
		\For {(all $\uz \in \Zq^t$ s.t. $\sigma(\cale,\uz) = s$  )}
        \State Decode $\{A(x,y;\uz) \mid (x,y) \in \cale \setminus \cale_{2,\underline{z}}\}$ and $\{B(x,y;\uz) \mid (x,y) \in \cale_{2,\underline{z}}\}$
        by invoking \textbf{S-MDS-DEC}($\uz, \{A(x,y;\uz') 
   \mid \uz' \text{ s.t. } \sigma(\cale, \uz) < s\}$)
		\EndFor
		\State Apply ${\cal L}^{-1}$ on $\{B(x,y;\uz) \mid (x,y) \in \cale_{2,\underline{z}}\}$ to get $\{A(x,y;\uz) \mid (x,y) \in \cale_{2,\underline{z}}\}$
		\State $s = s+1$
		\EndWhile
	\end{algorithmic}
\end{algorithm} 

\subsection{Systematic Encoding \label{sec:encoding}}

The $k$ systematic nodes contain the $k\alpha$ message symbols. It is required to encode these message symbols to obtain $(n-k)\alpha$ symbols belonging to the parity nodes. It can be done by invoking the sequential decoding algorithm by assuming that all the $(n-k)$ parity nodes are erased. Thus we make use of the decoding algorithm to carry out the systematic encoding as well. 

\subsection{An Example for $q=4,t=5$}

Let ${\cal E} = \{ (1,2), (2,2), (2,3), (3,3) \}$. In this case, $\sigma({\cal E},\uz)$ can take on values from $\{0,1,2\}$. Let us consider a plane $\underline{z}_1 = (1,0,1,1,0)$, represented by
\bean
\text{Pict}({\cal E},\underline{z}_1) \ = \ \left[ \begin{array}{ccccc} 0 & 1 & 0 & 0 & 1 \\ 1 & \circled{0} & 1 & 1 & 0 \\ 0 & \circled{0} & \circled{0} & 0 & 0 \\ 0 & 0 & \circled{0} & 0 & 0 \end{array} \right] .
\eean
in combination with the given erasure pattern. The intersection score $\sigma({\cal E}, \underline{z}_1) = 0$ in this case, and \eqref{eq:wh_eq_0} becomes
\bean
\left[ \begin{array}{cccc} \Theta_{(1,2)} & \Theta_{(2,2)} & \Theta_{(2,3)} & \Theta_{(3,3)} \end{array} \right] \left[ \begin{array}{c} A(1,2;\underline{z}_1)  \\
	A(2,2;\underline{z}_1) \\ 
	A(2,3;\underline{z}_1) \\ 
	A(3,3;\underline{z}_1) \end{array} \right] & = & \kappa_{*}
\eean
where the vector $\Theta_{(x,y)}$ is the $(x,y)$-th column of $\Theta$, for all $x \in \mathbb{Z}_q, \ y \in [t]$.
As any $q=4$ columns of $\Theta$ together form an invertible matrix, the erased symbols on $\underline{z}_1$ can be decoded. In the first round, planes with score value $1$ will be decoded assuming that the zeroth round is finished. As the principle remains the same, we skip the case of score equal to $1$, and directly proceed to considering a plane $\underline{z}_2$ with score $\sigma({\cal E}, \underline{z}_2) = 2$. Let $\underline{z}_2 = (1,2,3,1,0)$ represented by
\bean
\text{Pict}({\cal E},\underline{z}_2) \ = \ \left[ \begin{array}{ccccc} 0 & 0 & 0 & 0 & 1 \\ 1 & \circled{0} & 0 & 1 & 0 \\ 0 & {\color{red}\circled{1}} & \circled{0} & 0 & 0 \\ 0 & 0 & {\color{red}\circled{1}} & 0 & 0 \end{array} \right] .
\eean
For the plane $\underline{z}_2$, we have ${\cal E}_{0,\underline{z}_2} \ = \ \{(2,2),(3,3) \}, \ {\cal E}_{1,\underline{z}_2} \ = \ \phi, \ \text{and} \ {\cal E}_{2,\underline{z}_2} \ = \ \{ (1,2), (2,3) \}$. Thus the equation \eqref{eq:focused_pc_eqn_v4}, obtained after substituting symbols recovered in zeroth and first rounds, takes the form
\bean
\left[ \begin{array}{cccc} \Theta_{(1,2)} & \Theta_{(2,2)} & \Theta_{(2,3)} & \Theta_{(3,3)} \end{array} \right] \left[ \begin{array}{c} B(1,2;\underline{z}_2)  \\
	A(2,2;\underline{z}_2) \\ 
	B(2,3;\underline{z}_2) \\ 
	A(3,3;\underline{z}_2) \end{array} \right] & = & \kappa_{*}.
\eean
Once these unknown symbols are decoded, it remains to recover symbol $A(1,2;\underline{z}_2)$ (say for instance) from $B(1,2;\underline{z}_2)$. One can observe that the companion term of $B(1,2;\underline{z}_2)$ belongs to the plane $\underline{z}_3 = (1,1,3,1,0)$ represented by
\bean
\text{Pict}({\cal E},\underline{z}_3) \ = \ \left[ \begin{array}{ccccc} 0 & 0 & 0 & 0 & 1 \\ 1 & {\color{red}\circled{1}} & 0 & 1 & 0 \\ 0 & \circled{0} & \circled{0} & 0 & 0 \\ 0 & 0 & {\color{red}\circled{1}} & 0 & 0 \end{array} \right] .
\eean
The plane $\underline{z}_3$ has the same score $2$, and therefore the companion term $B(2,2;\underline{z}_3)$ has  been decoded in the same round. We apply the transformation ${\cal L}^{-1}$ on vector $[B(1,2;\underline{z}_2) \ B(2,2;\underline{z}_3)]^T$ to recover back $A(1,2;\underline{z}_2)$. This completes the decoding of all erased symbols in $\underline{z}_2$. 

\subsection{Comparison with the Recent Results of Ye and Barg \cite{YeBar_2} \label{sec:comparison}}

The parameter set 
\bean
n \ = \ qt, k \ = \ q(t-1), d \ = \ (n - 1), \alpha \ = \ q^t , \beta \ = \  q^{t-1} 
\eean
was first introduced in \cite{SasAgaKum} by Sasidharan et. al. and is common to the MSR codes introduced here as well as in \cite{YeBar_2}.  In terms of the transformed code symbols (see Sec.~\ref{sec:transform}), the parity-check equations of the coupled-layer MSR code correspond to that of $q^t$ independent scalar MDS codes:
	\beq \label{eq:bcode_1}
	\sum\limits_{x \in \mathbb{Z}_q} \sum\limits_{y \in [t]}   \tlxy \bxyzexp =  0, \ \underline{z} \in \mathbb{Z}_q^t, \ \ell \in [0,q-1].  
	\eeq
For ease of reference, we will refer to the collection of symbols $\{ B(x,y;\underline{z}) \mid x \in \mathbb{Z}_q, y \in [t], \underline{z} \in \mathbb{Z}_q^t\}$ forming the decoupled code ${\cal D}$, as the $B$-code. The transformation ${\cal L}$ that takes the code symbols \axyz\ forming the the original code ${\cal C}$ to the $B$-code is given in \eqref{eq:bcode}.   

Since the presentations of the codes in \cite{YeBar_2} and the current paper are quite different, to make the connection between the constructions presented in the two papers, it will be found convenient to associate an analogous $B$-code for the code constructed\footnote{Such an association is not however, a part of the presentation in \cite{YeBar_2}.} in \cite{YeBar_2} as well.  Let $B_{\text{\scriptsize YB}}(x,y;\underline{z})$ denote the transformed code symbols in the construction of \cite{YeBar_2}. Then the linear transformation ${\cal L}_{\text{\scriptsize YB}}$ is described below: 
	\beq
	\left[ \begin{array}{c} B_{\text{\scriptsize YB}}(x,y;z_y,\underline{z}_{\sim y}) \\ B_{\text{\scriptsize YB}}(z_y,y;x,\underline{z}_{\sim y}) \end{array} \right]  =  
	\left[ \begin{array}{cc} 1 &  1 \\ u & 1 \end{array} \right] 
	\left[ \begin{array}{c} 
		\axyzexp \\ \axyzsw
	\end{array} \right].  \label{eq:transform_1}
	\eeq 
In setting down this transformation, one must ensure that in the ordering of elements within each of the two column vectors on either side of the transformation is such that $B_{\text{\scriptsize YB}}(x,y;z_y,\underline{z}_{\sim y})$ appears on top of $B_{\text{\scriptsize YB}}(z_y,y;x,\underline{z}_{\sim y})$ provided $x > z_y$. Similarly with the $A(\cdot)$'s.    When viewed from this angle, the two constructions differ only in the $(2 \times 2)$ linear transformation used to pass from coupled to decoupled code.  Similarities between the two constructions were not apparent to the authors at the time of the initial submission. 

\section{A More Generic Description of the Code} \label{sec:generic} 

In this section, we present a more generic description of the MSR code in which 
\bit
\item the scalar MDS code discussed above is replaced by an MDS code over an arbitrary alphabet \calq\ of size $Q \geq n$, for example, by a binary MDS codes, i.e., a code that is MDS, but over a binary-vector alphabet, i.e., an alphabet of the form $\mathbb{F}_2^{m}$, some $m \geq 2$
\item the coupling transformation is replaced by a symbol mapping coming from a second MDS code of length $4$ with $2$ data symbols.
\eit
As before, there are three ingredients to the construction.  
\ben
\item {\em Numerology:} \  The parameters of the MSR code to be constructed remain as:
\bean
\left\{ (n, k, d), \ \ (\alpha, \beta)\right\}.
\eean
However, the alphabet is this time, a generic alphabet \calq\ of size $|\calq|=Q$.  By numerology, we mean here, the selection:
\bean
n=qt, \ q,t \geq 2, \ \ \text{ $k$ determined from $(n-k)=q$, \ \ $\alpha=q^t$ \ and \ $d=(n-1)$}.  
\eean
With this, we get $\beta=\frac{\alpha}{(d-k+1)} \ = \ q^{(t-1)}$.  This choice of parameters was first made in \cite{SasAgaKum}.  
\item {\em Constituent MDS Codes:  } 
The construction makes use of an MDS code \cmds\ over an alphabet \calq\ having parameters 
\bean
\text{ block length} = n, \ \  \text{ size} = \ |\calq| ^k =Q^k, \ \text{minimum distance } d_{\min}=(n-k+1) .
\eean
Thus this code can recover from any pattern of $(n-k)$ erasures and this is the only property that we will require of the MDS code.

\item {\em Layering, Symbol-Pairing and Coupling of MDS Codes:  } 
The code \calc\ is easiest described in terms of a $3$-step encoding process.  This description does not result in a systematic code and hence an alternative procedure, described later in this section, may be employed in practice.  
\ben[(a)]
\item {\em Layering:} \ In the first step, a collection of $q^t$ data sets, with $k$ data symbols from \calq\ contained in each data set are formed.  Each $k$-set is then encoded using the MDS code \cmds\ into a collection of $q^t$ codewords, drawn from \cmds. The $q^t$ codewords are organized into layers in which the layers are indexed by a parameter $\uz \ = \ [z_1,z_2,\cdots,z_t]^t$, $\uz \in \mathbb{Z}_q^t$.  The code symbols within a layer are indexed by a pair of coordinates $(x,y), \xyin$.  The $(x,y)$th symbol in the \uz\ th layer is noted by $B(x,y;\uz)$.  
\item {\em Symbol Pairing:} \ In the second step, symbols from the $B$ code are paired up:
\bean
\bxyz\ \text{ is paired with } \bxyzsw \text{ whenever $x \neq z_y$ }
\eean
The two symbols will be referred to as companions. The symbols
\bean
\bxyz\ \ \ x =z_y, 
\eean
remain unpaired and will be referred to as fixed points for reasons that will become clear shortly. 
\item {\em Coupling:} \ In the third and final step, the symbols \bxyz\ and \bxyzsw\ are mapped onto a second pair \axyz\ and \axyzsw\ of symbols in such a way that the $4$-tuple 
\bean
\left( \bxyz, \bxyzsw, \axyz, \axyzsw \right)  
\eean
is always a codewords of a fixed $(4,Q^2,3)$ MDS code.  In particular, all four of these symbols can be recovered just from knowing any $2$ of these symbols.  For the cases when $x = z_y$, we set 
\bean
\axyz & = & \bxyz .
\eean
The nodes are indexed by $(x,y)$ with \xyin.  The $(x,y)$th node then stores the symbols \axyz.  With this, the description of the code is complete. 
\een

\subsection{Data Collection}
Data collection can be seen as a process of recovering from a set of $(n-k)=q$ erasures.  One first recovers data from planes \uz\ having lower intersection scores before proceeding to decide layers with larger intersection score.  The intersection score of  plane \uz\ is the number of fixed points in the plane that have been erased, i.e., the number of code symbols \axyz\ in the plane \uz\ with $(x,y) \in {\cal E}$.  
We continue to partition the erased symbols into three classes.  Given an erasure pattern \cale\ and a plane \uz\ we have 
\bean
& \cale_{0,\underline{z}} \ = \ \left\{ (x,y) \in \cale \mid x=z_y \right\}, & \\ & \text{(fixed points in the plane that have been erased)} & \\
& \cale_{1,\underline{z}} \ = \  \left\{ (x,y) \in \cale \mid (z_y,y) \notin \cale \text{ hence $x \neq z_y$} \right\}, & \\
& \text{(coordinates of erased symbols in plane \uz\ whose companions have not been erased)} & \\
& \cale_{2,\underline{z}} \ = \  \left\{ (x,y) \in \cale \mid (z_y,y) \in \cale, \ x \neq z_y \right\}  & \\ 
&  \text{(coordinates of erased symbols in plane \uz\ whose companions have also been erased)}. & 
\eean
The key to decoding in sequential fashion is to recognize that 
\ben
\item {\em Handling Non-Erased Symbols:} \ \ In any plane, the companion of an non-erased symbol is either a non-erased symbol or else, an erased symbol belonging to a plane having a lower intersection score.  Under the sequential decoding procedure adopted here, one may assume that symbols in planes having a lower intersection score have already been decoded.  Hence we may assume here as well, that the companion of a non-erased symbol is also, a non-erased symbol.  Hence, in the case of a non-erased symbol \axyz, we may assume that both \axyz\ and \axyzsw\ are known and hence \bxyz\ can be computed since the $4$-tuple $\left( \bxyz, \bxyzsw, \axyz, \axyzsw \right)$ is an MDS code with block length $4$ that can be recovered from knowledge of any two symbols.  
\item {\em Handling Erased Symbols Lying in $\cale_{1,\underline{z}}$:} \ \ The companion of an erased symbol belonging to set 
\bean
\cale_{1,\underline{z}} & = & \left\{ (x,y) \in \cale \mid (z_y,y) \notin \cale \text{ hence $x \neq z_y$} \right\}, 
\eean
is a non-erased symbol and hence if \bxyz\ can be recovered through decoding of the MDS code in that plane, one can recover \axyz from the pair \bxyz\ and \axyzsw.
\item {\em Handling Erased Symbols Lying in $\cale_{0,\underline{z}}$:} \ \ In the case of erased symbols belonging to set 
\bean
\cale_{0,\underline{z}} & = & \left\{ (x,y) \in \cale \mid x=z_y \right\}
\eean
we have $\axyz \ = \ \bxyz$.  Hence if we solve for \bxyz\, we will have recovered \axyz\ as well.
\item {\em Handling Erased Symbols Lying in $\cale_{2,\underline{z}}$:} \ \ Finally, in the case of erased symbols belonging to the set 
\bean
\cale_{2,\underline{z}} & = & \left\{ (x,y) \in \cale \mid (z_y,y) \in \cale, \ x \neq z_y \right\}   , 
\eean
we have that the companion of \bxyz\ belongs to a plane having the same intersection score and hence is decoded in the same phase.  Given \bxyz\ and \bxyzsw\, we can recover \axyz. 
\een
It follows from this that systematic encoding can be accomplished by filling in data symbols into $k$ systematic nodes and recovering the remaining $(n-k)=q$ symbols through decoding. 
\een

\subsection{Node Repair}

Let us assume that node $(x_0,y_0)$ is the node to be repaired. Let 
\bea \label{eq:rd_planes}
{\cal Z}_0 & = & \left\{ \uz \mid z_{y_0} = x_0     \right\}, 
\eea 
be the collection of planes having $(x_0,y_0)$ as a fixed point.  The number of such planes is clearly equal to $q^{t-1}$.  Repair then proceeds as follows: 
\ben
\item Each remaining node $(x,y)\neq (x_0,y_0)$, passes on the $\beta=q^{t-1}$ symbols 
\bean
\left\{ \axyz \mid \uz \in {\cal Z}_0 \right\} ,
\eean
contained within that node, to the replacement node.  
\item Consider the collection of symbols $\{A(x,y;\uz) \mid \xyin \}$ for fixed $\uz \in \zo$:  
\ben[(a)]
\item For $y \neq y_0$, it can be verified that the companion \axyzsw\ of \axyz\ is a non-erased symbol.  Hence for such symbols, the corresponding value of \bxyz\ is known.   
\item This leaves us with at most $q$ unknown values of $\{ \bxyz \mid x \in \mathbb{Z}_q, y=y_0 \}$ in the plane \uz.  These can be decoded by making the use of the MDS code formed by the symbols $\{ \bxyz \mid \xyin \}$. 
\item In the case of the symbol $A(x_0,y_0;\uz)$, this symbol is a fixed point meaning that $A(x_0,y_0;\uz)=B(x_0,y_0;\uz)$ and hence once the MDS code corresponding to the plane \uz\ has been decoded, we know the value of $B(x_0,y_0;\uz)$ and hence that of $A(x_0,y_0;\uz)$.    
\item In the case of the symbols $\{\axyz \mid x \neq x_0, y=y_0\}$, the symbols \axyz\ are non-erased and the symbol values \bxyz\ have been determined.  Hence both \axyz\ and \bxyz\ are known. The companions of the $q^t$ symbols $\{\axyz \mid x \in \mathbb{Z}_q, y=y_0\}$ are precisely the $q^t$ symbols continued in the erased node.  Since we know both \axyz\ as well as \bxyz\ in these instances, we can then recover the values of the companion terms \axyzsw.   
\een 
This completes description of the recovery process.
\een

A tabular listing of the parameters of codes where a binary MDS code can be employed appears in Table~\ref{tab:bin_mds}. To obtain the level of sub-packetization over the binary field, as opposed to over the alphabet \calq, one simply multiples the values of both $\alpha$ and $\beta$ by $m$. 

	{\scriptsize
\begin{table}
\caption{Parameters of MSR codes constructed using Binary MDS codes as building blocks.}
	\bc
	$(n,k,d=n-1,\alpha, \beta)$-MSR Codes constructed from various $(n,k)$-MDS codes over a vector alphabet of  $m$-tuples. 
	\ec
	\bean
	\begin{array}{|c|c|c|c|c|c|c|}
		\hline \hline
		\text{Base MDS code} & m & n & k  & \alpha & \beta & \text{Field size} \\ \hline \hline						\text{Reed-Solomon} & 1 & qt & q(t-1) & q^t & q^{t-1} & n \\ \hline
		\text{RDP} & (p-1) & (p+1) & (p-1) & 2^{\frac{p+1}{2}} & 2^{\frac{p-1}{2}} & \text{Binary} \\ 
		 & p \text{ odd prime} &  &  &   &  &  \\ \hline
		\text{$(6,4)$-RDP} & 4 & 6 & 4 & 8 &  4 & \text{Binary} \\ \hline
		\text{Evenodd} & (p-1) & (p+2) & (p-1) & 3^{\frac{p+2}{3}} & 3^{\frac{p-1}{3}} & \text{Binary} \\ 
		 \text{(generalized)} & p \text{ prime} &  &  &   &  &  \\ 
		  & 3 \mid (p-1)   &  &  &   &  &  \\ \hline
		\text{$(7,4)$-Evenodd} & 6 & 9 & 6 &  27 &  9 & \text{Binary} \\ \hline
	\end{array}
	\eean  \label{tab:bin_mds}
\end{table}
	 }

 \section{Node Repair for $d<(n-1)$ Under Restricted Helper Node Sets} \label{sec:smaller_d} 
 
 In this section we note that the construction can be extended to handle the repair of a failed node for the range $q \leq d \leq n-2$, and $k \leq d$.  Let us set 
 \bean
 n-1-d & = & a.
 \eean
 Hence $a$ represents the number of nodes that do not participate in the repair process.  We will term node that does not participate in the repair process as an {\em aloof} node, hence there are $a$ aloof nodes. We considered above the case when $a=0$.  Our aim here is to show how one can extend the construction to the case when $1 \leq a \leq n-q-1$.
 
We will maintain the value of $d-k+1=q$, $\alpha=q^t$, $\beta=q^{t-1}$.  Hence with $d=n-1-a$, we have that $n-1-a-k+1=n-k-a=q$, so that 
 \bean
 n-k & = & q+a \\
 k & = & n-q-a.
 \eean
  Hence we replace the earlier $(n,Q^{n-q},q+1)$ MDS codes with MDS codes having parameters $(n,Q^{n-q-a},q+a+1)$.
 
We will illustrate below with the cases $d=n-2$. The general case follows along similar lines.  Given a failed node $(x_0,y_0)$ we choose the symbols from the planes \zo\ as the symbols transmitted by the helper nodes.  
 
\subsection{Case $d=n-2$}

Since $d-k+1=q$, we have that $n-k=n-d+q-1=q+1$.  Hence the MDS codes in each plane have parameters $(n,Q^{n-q-1},q+2)$.  this MDS code is capable of recovering from $(q+1)$ erasures. 

We restrict our attention on planes \uz\ such that $\uz \in \zo$, where $\zo$ is as defined in \eqref{eq:rd_planes}.   For the case when $d=(n-2)$, we have that there is a single aloof node.  We calculate an intersection score for each plane in \zo.  Clearly since $(x_0,y_0)$ has been erased (it is the failed node) and we are dealing with planes in \zo, the smallest possible intersection score equals $1$. 

 {\em Helper Node Restriction:}  We will assume that the helper nodes include all nodes lying in the same ``$y$-section'',  i.e., all include the $(q-1)$ nodes corresponding to 
\bean
\left\{ (x,y) \mid x \in \mathbb{Z}_q, x \ne x_0, \ y =y_0\right\}.
\eean

We handle repair of the symbols of the failed node by considering the planes within \zo\ in increasing order of intersection score.   Let the aloof node have coordinates $(x_a,y_a)$.

Let \uz\ be a plane in \zo\ having intersection score equals $1$.    In such planes, the aloof node is not a fixed point.  Hence the number of unknown symbols \bxyz\ in such planes is $(q+1)$ and these can be recovered from the properties of the MDS code.  Additionally, the value $A(x_a,y_a;\uz)$ can be recovered as the companion of $B(x_a,y_a;\uz)$ though it is a non-erased symbol.  On these $(q+1)$ symbols are recovered, repair proceeds as before. 

If the plane \uz\ is such that the intersection score equals $2$, then the aloof node is a fixed point.   The aloof node then can potentially result in some additional values of \bxyz\ being unknown in this plane corresponding to $(x,y_a), x\neq x_a$.  However, for each such coordinates, the companion lies in a plane with lesser intersection score.  Hence for such symbols we can compute \bxyz\ and hence once again, there are only $(q+1)$ unknown $B$-symbols in the plane which can be recovered using the properties of the MDS code.  
 
		\begin{figure}[h!]
			\begin{center}
				\subfigure[A plane with intersection score $0$ for the case $q=4, t=5, d=n-2$. The aloof node is $A2$ here. ]{\label{fig:aloof1}\includegraphics[width=2.5in]{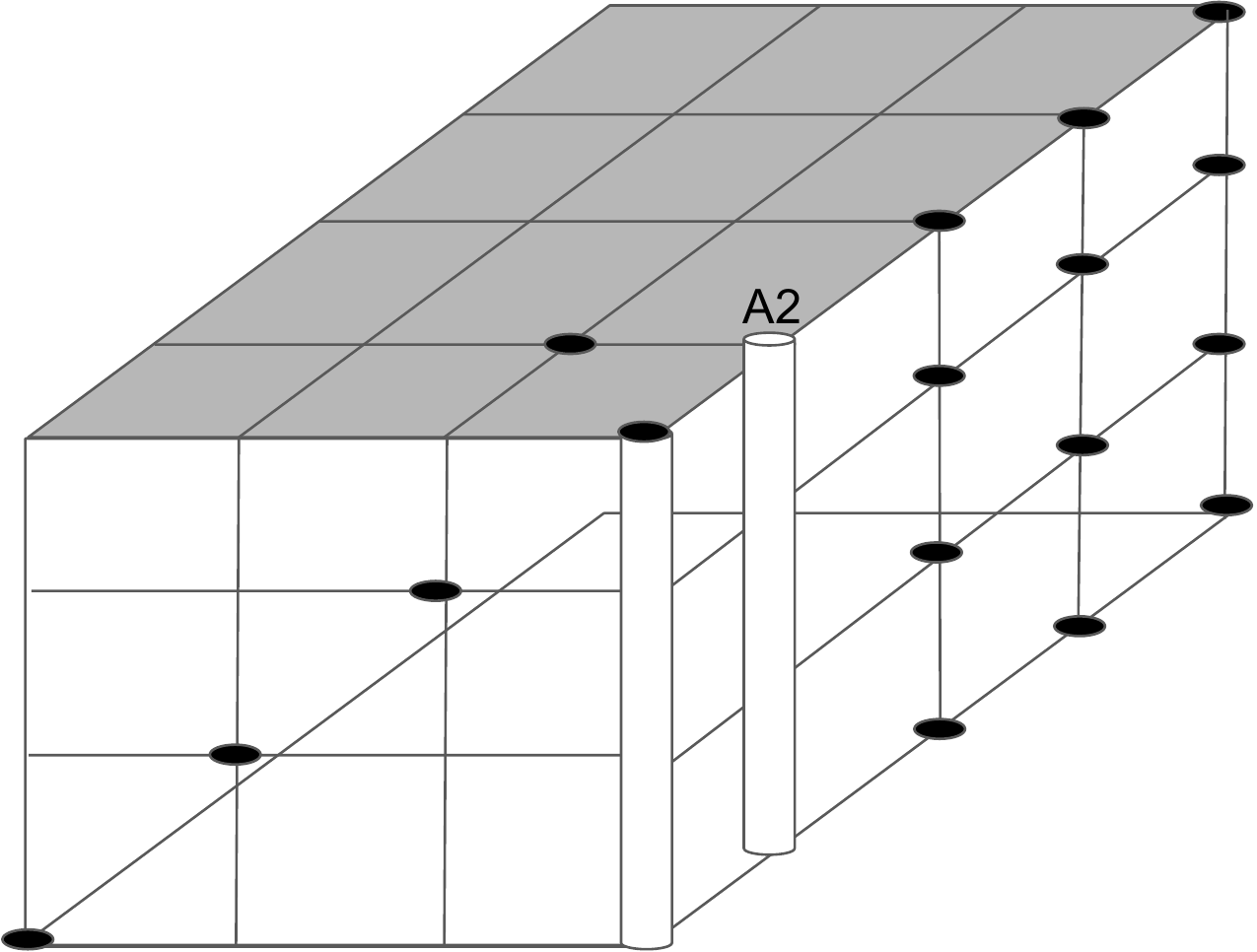}}
				\hspace{0.2in}
				\subfigure[A plane with intersection score $1$ for the case $q=4, t=5, d=n-2$.]{\label{fig:aloof2}\includegraphics[width=2.5in]{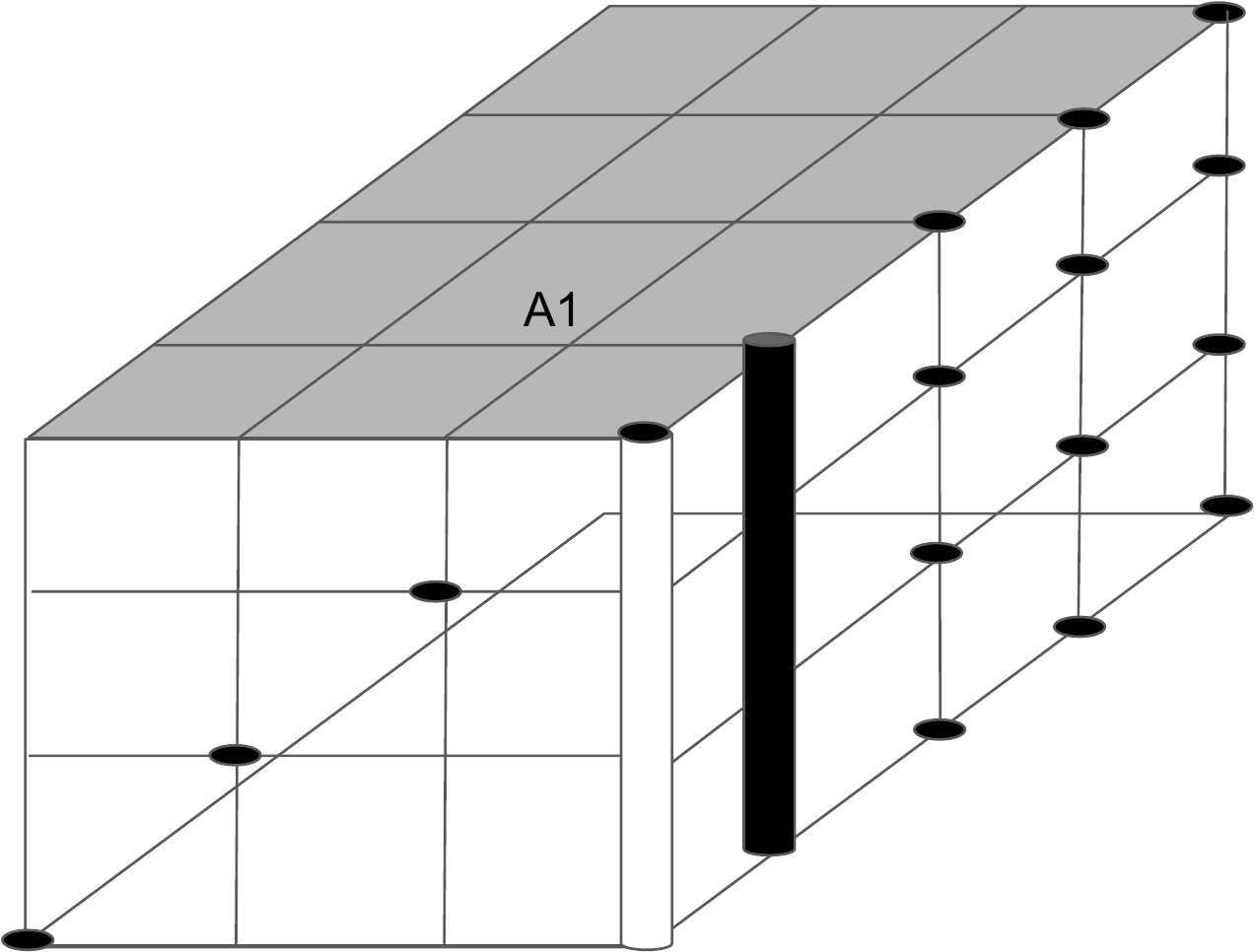}}
				\caption{Illustration of the repair planes for $d=n-2$.\label{fig:repair2}}
			\end{center}
		\end{figure}

\bibliographystyle{IEEEtran}
\bibliography{netcod}

\end{document}